\documentstyle[12pt]{article}
\input epsf

\begin{document}

\title{Radiative Correction to the Transferred
Polarization in Elastic
Electron-Proton Scattering}

\author{A.V. Afanasev$^{a)}$\footnote{On leave
of absence from Kharkov Institute of Physics and Technology 63108, 
Kharkov, Ukraine}, 
I. Akushevich$^{a)}$\footnote{On leave
of absence from National
Center of Particle and High Energy Physics,
220040 Minsk, Belarus}, N.P.Merenkov$^{b)}$ }
\date{}
\maketitle
\begin{center}
{\small {\it $^{(a)}$ North Carolina Central University,
Durham, NC 27707, USA \\ and  \\
TJNAF, Newport News, VA 23606, USA\\}}
{\small {\it{$^{(b)}$ NSC "Kharkov Institute of Physics and Technology" \\
}}}
{\small {\it {63108, Akademicheskaya 1, Kharkov, Ukraine}}}
\end{center}
\begin{abstract}

Model independent radiative correction to 
the recoil proton polarization for the elastic electron--proton
scattering is calculated within method of electron structure functions. The
explicit expressions for the recoil proton polarization are
represented as a contraction of the electron structure and
the hard part of the polarization dependent contribution into
cross--section. The calculation of the hard part with first order
radiative correction is performed. The obtained representation includes
the leading radiative corrections in all orders of perturbation theory and
the main part of the second order next--to--leading ones. Numerical
calculations illustrate our analytical results.
\end{abstract}

\section{Introduction}

\hspace{0.6cm}
It was proposed over 25 years ago \cite{sombodyMPKarlson} that recoil proton
polarization in the elastic process $\vec e+p \to e+\vec P$, can be used to
measure the proton electric form factor ($G_{EP}$). This method provides an
alternative to the Rosenbluth separation and appears to be more sensitive to
$G_{EP}$ in the GeV--range of 4--momentum transfers ($Q^2$). 
Such measurements were done
first at MIT-Bates \cite{Bates} and later on extended to higher 
$Q^2=3.5$ GeV$^2$ at
Jefferson Lab \cite{1CEBAFothers}. The latter experiment provided 
the first evidence
of significant deviation of $G_{EP}$ from the dipole form at higher $Q^2$.

In the recent Jefferson Lab experiment  \cite{1CEBAFothers}  the
events	corresponding to elastic process
\begin{equation}\label{1}
\vec{e}^{\,-}(k_1) + P(p_1) \rightarrow e^-(k_2) + \vec{P}(p_2)
\end{equation}
as well as radiative process
\begin{equation}\label{2}
\vec{e}^{\,-}(k_1) + P(p_1) \rightarrow e^-(k_2) + \gamma(k) + \vec{P}(p_2)
\end{equation}
have been analyzed.

The main goal of these experiments is the measurement of the proton
electric formfactor $G_E.$ It can be done because the ratio of the
longitudinal polarization of recoil proton to the transverse one in
Born approximation is proportional to the ratio $G_M/G_E$
 \cite{sombodyMPKarlson} where $G_M$ is
the well known proton magnetic formfactor.
This statement is valid if 3--vector of the longitudinal polarization has
orientation along the recoil proton 3--momentum, and 3--vector of the
transverse polarization is within  the plane $(\vec k_1, \vec p_2).$
The interpretation of these high-precision experiments in terms of the
proton electromagnetic formfactors $G_M$ and $G_E$ requires  adequate
theoretical calculations with a per cent accuracy or
better. Such  calculations must include  the
first order radiative corrections (RC) to the elastic cross--section ( due
to radiation of real soft and virtual photon) and full analysis of the
radiative events. Moreover, leading higher order corrections have to be
taken into account.

All the corresponding contributions can be joint within the framework of
the electron structure function representation, which is a QED analog of
the
well known Drell--Yan representation \cite{DrellYan}. This representation
was applied before for the calculation of the RC to unpolarized
electron--positron annihilation \cite{KuraevFadin} and deep inelastic
scattering \cite{KurMerFad} cross--sections.

In the present work we
generalize the electron structure function representation for the case of
 scattering of  polarized particles, namely for the analysis of
the recoil proton polarization in elastic ep-scattering.

\section{The leading approximation}

\hspace{0.6cm}

The cross--section of the quasireal electron--proton scattering in the
framework of the electron structure function method can be written as a
contraction of two electron structure functions, that corresponds to the
possibility to radiate hard collinear as well as virtual and soft photons
and electron--positron pairs by both the initial and the scattered
electron, and hard part of the cross--section that depends on shifted
4--momenta. This representation follows from the quasireal electron method
\cite{BaierFadinkhoze} that is suitable for description of the
collinear radiation.

In the problem considered here we will be interested in the spin
dependent part of the cross--section only. For this case the
corresponding representation can be written as
\begin{equation}\label{3}
\frac{d\sigma^{\parallel,\bot}(k_1,k_2)}{d\,Q^2d\,y} =
\int\limits_{z_{1m}}^1d\,z_1\int \limits_{z_{2m}}^1d\,z_2
D^{(p)}(z_1,L)\frac{1}{z_2^2} D^{(u)}(z_2,L)
\frac{d\sigma^{\parallel,\bot(hard)}(\hat
k_1,\hat k_2)}{d\,\hat{Q}^2\,d\,\hat{y}} \ , \ \ L=\ln\frac{Q^2}{m^2}\ ,
\end{equation} where $m$ is the electron mass,
\begin{equation}\label{4}
\hat k_1 = z_1k_1\ , \ \hat k_2 =\frac{k_2}{z_2}\ , \ Q^2 = -(k_1-k_2)^2 \
, \ \hat{Q}^2 = -(\hat k_1-\hat k_2)^2 = \frac{z_1}{z_2}Q^2\ ,
\end{equation}
$$y=\frac{2p_1(k_1-k_2)}{V}\ , \ \hat{y} = 1-\frac{1-y}{z_1z_2}\ , \
V=2p_1k_1 \ . $$

The electron structure function $D^{(p)}(z_1,L)$ is responsible for
radiation by the initial polarized electron, whereas the function
$D^{(u)}(z_2,L)$ describes   radiation by the scattered unpolarized
electron. The photonic contribution into the electron structure function
is the same for polarized and unpolarized cases, but the contribution
due to pair production differs in the singlet channel \cite{KonchMerShekh}.
Therefore we can write
\begin{equation}\label{5}
D^{(u)}(z,L) = D^{\gamma}(z,L) + D^{^{e^+e^-}}_N +D^{^{e^+e^-}(u)}_S  \ ,
\end{equation}
\begin{equation}\label{6}
D^{(p)}(z,L) = D^{\gamma}(z,L) + D^{^{e^+e^-}}_N +D^{^{e^+e^-}(p)}_S  \ .
\end{equation}
There exists many different representations for the photonic contribution
into the structure function \cite{strucfunc}, but here we will use the
form given in \cite{KuraevFadin} for $D^{\gamma}$, $D^{^{e^+e^-}}_N$ and
$D^{^{e^+e^-}(u)}_S$
\begin{equation}\label{7}
D^{\gamma}(z,Q^2) =
\frac{1}{2}\beta(1-z)^{\beta/2-1}\Bigl[1+\frac{3}{8}\beta
-\frac{\beta^2}{48}\bigl(\frac{1}{3}L+\pi^2-\frac{47}{8}\bigr)\Bigr]-\frac
{\beta}{4}(1+z) +
\end{equation}
$$\frac{\beta^2}{32}\Big[-4(1+z)\ln(1-z)-\frac{1+3z^2}{1-z}\ln\,z-5-z\Bigr]
\ , \ \ \beta = \frac{2\alpha}{\pi}(L-1)\ . $$
\begin{equation}\label{10,pair contribution into EST}
D^{^{e^+e^-}}_N(z,Q^2) =
\frac{\alpha^2}{\pi^2}\Bigl[\frac{1}{12(1-z)}\bigl(1-z-\frac{2m}{\varepsilon}
\bigr)^{\beta/2}\bigl(L_1-\frac{5}{3}\bigr)^2\bigl(1+z^2+\frac{\beta}{6}
\bigl(L_1-\frac{5}{3}\bigr)\bigr)\Bigr]
\theta\bigl(1-z-\frac{2m}{\varepsilon}\bigr) \ ,
\end{equation}
\begin{equation}
D^{^{e^+e^-}(u)}_S = \frac{\alpha^2}{4\pi^2}L^2\bigl[\frac{2(1-z^3)}{3z}
+\frac{1}{2}(1-z)
+(1+z)\ln{z}\bigr]\theta\bigl(1-z-\frac{2m}{\varepsilon}\bigr) \ ,
\end{equation}
\begin{equation}
D^{^{e^+e^-}(p)}_S=\frac{\alpha^2}{4\pi^2}L^2\bigl(\frac{5(1-z)}{2}+(1+z)
\ln\,z\bigr)\theta\bigl(1-z-\frac{2m}{\varepsilon}\bigr)\ ,
\end{equation}
where $\varepsilon$ is the energy of the parent electron and $L_1= L +
2\ln(1-z).$
The above form of the structure function $D^{^{e^+e^-}}_N$ includes
effects due to real pair production only. The correction caused by virtual
pair is included  in $D^{^{\gamma}}$. Note that the terms containing
$\alpha^2L^3$
cancel each other in the sum $D^{^{\gamma}} + D^{^{e^+e^-}}_N.$

Instead of the photonic structure function given by Eq.~(7), one can use
the its iterative form \cite{Jad}
\begin{equation}\label{11, iterative form EST}
D^{\gamma}(z,L) = \delta(1-z) +
\sum_{k=1}^{\infty}\frac{1}{k!}
\biggl(\frac{\alpha\,L}{2\pi}\biggr)^kP_1(z)^{\otimes k}\ ,
\end{equation}
$$\underbrace{ P_1(z)\otimes\cdots\otimes P_1(z)}_{k} = P_1(z)^{\otimes k}
\ , \quad P_1(z)\otimes P_1(z) = \int\limits_{
z}^{1}P_1(t)P_1\biggl(\frac{z}{t} \biggr)\frac{dt}{t} \ ,$$
$$P_1(z) = \frac{1+z^2}{1-z}\theta(1-z-\Delta) +
\delta(1-z)\bigl(2\ln{\Delta}+\frac{3}{2}\bigr) \ , \ \Delta \ll 1 \ . $$
The iterative form (11) of $D^{^{\gamma}}$ does not include any
effects caused by pair production. The corresponding
nonsinglet part of the structure due to real and virtual pair production
can be inserted into iterative form of
$D^{\gamma}(z,L)$ by replacing $\alpha L/2\pi$ on the right
side of Eq.~(11) by the effective electromagnetic coupling
\begin{equation}\label{effective coupling} \frac{\alpha_{eff}}{2\pi} =
-\frac{3}{2}\ln{\bigl(1-\frac{\alpha L}{3\pi}\bigr)}, \end{equation}
which is the integral of the running electromagnetic constant.

The limits of integration with respect to $z_1$ and $z_2$ in the master
formula
(3) can be found from the constraint on the Bjorken variable $\hat x$ for
the partonic process
\begin{equation}\label{13}
\hat x = \frac{-(\hat{k_1}-\hat{k_2})^2}{2p_1(\hat{k_1}-\hat{k_2})} =
\frac{z_1yx}{z_1z_2 +y-1} < 1 \ , \ \ x=\frac{Q^2}{2p_1(k_1-k_2)}\ .
\end{equation}
By taking into account also that $z_{1,2}<1$ and $xy=Q^2/V$, we derive
from (13)
\begin{equation}\label{14,limits}
1>z_2>z_{2m} \ , \ 1>z_1>z_{1m} \ , \
z_{2m}=\frac{1-y}{z_1}+\frac{Q^2}{V}\ , \ z_{1m}=\frac{V(1-y)}{V-Q^2}\ .
\end{equation}

In the framework of the leading logarithmic approximation we have to
take the
elastic (Born) cross--section as the hard part under the integral
on the right hand side of Eq.~(3)
\begin{equation}\label{15, leading}
\frac{d\sigma^{^{\parallel,\bot(B)}}_{hard}}{d\,Q^2d\,y}=
\frac{d\sigma^{^{\parallel,\bot(B)}}}{d\,Q^2}\delta\bigl(y-\frac
{Q^2}{V}\bigr)\ .
\end{equation}

In the case of the longitudinal polarization of the recoil proton, we
have
\begin{equation}\label{16} 
\frac{d\sigma^{^{\parallel(B)}}_{hard}}{d\,Q^2} =
\frac{4\pi\alpha^2(-Q^2)}{VQ^2}\bigl(1-\frac{Q^2}{2V}\bigr)\sqrt{\frac{Q^2}
{4M^2+Q^2}}G_M^2(-Q^2) \ .
\end{equation}
The quantity $\alpha(-Q^2)$ on the right hand side of Eq.~(16) is the
running electromagnetic constant that account for the effects of the vacuum
polarization
$$\alpha(q^2) = \frac{\alpha}{1-\frac{\alpha}{3\pi}
\ln{\frac{-q^2}{m^2}}} \ .$$

For the transverse polarization of the recoil proton, the hard
part of the cross--section reads
\begin{equation}\label{17} 
\frac{d\sigma^{^{\bot(B)}}_{hard}}{d\,Q^2} =
-2\frac{4\pi\alpha^2(-Q^2)}{VQ^2}{\frac{M}{\sqrt{Q^2+4M^2}}}\sqrt{1-\frac{Q^2}
{V}(1+\tau)}G_E(-Q^2)G_M(-Q^2)\ , \ \ \tau = \frac{M^2}{V}\ .
\end{equation}

Note that in zeroth order of perturbation theory the photonic
contribution into electron structure function gives an ordinary
$\delta$--function because (see also the iterative form (11))
\begin{equation}\label{18}
\lim_{\beta\rightarrow 0} \
\frac{1}{2}\beta(1-z)^{^{\frac{1}{2}\beta-1}} =
\delta(1-z) \ .
\end{equation}
It is easy to see that the representation (3) reproduces
the Born cross--section in this case
\begin{equation}\label{19}
\frac{d\sigma^{^{\parallel,\bot}}}{d\,Q^2d\,y}=\int\,dz_1\int\,dz_2\frac
{1}{z_2^2}\delta(1-z_1)\delta(1-z_2)\frac{d\sigma^{^{\parallel,\bot(B)}}}
{d\hat{Q}^2}\delta\bigl(\hat y-\frac{\hat Q^2}{\hat V}\bigr) =
\frac{d\sigma^{^{\parallel,\bot}(B)}}{d\,Q^2}\delta\bigl(y-\frac
{q^2}{V}\bigr).
\end{equation}

\section{Beyond the leading approximation}

\hspace{0.6cm}

We can improve the leading approximation for $d\sigma^{^{\parallel,\bot}}/
dQ^2dy $ given by formula (3) with $d\sigma^{^{\parallel,\bot(B)}}/dQ^2dy $
as a hard part of the cross--section under the integral. It can be
done by making more precise the expression namely for this hard part
\begin{equation}\label{20}
\frac{d\sigma^{^{\parallel,\bot}}_{hard}}{d\,Q^2d\,y} =
\frac{d\sigma^{^{\parallel,\bot(B)}}}{d\,Q^2d\,y} +
\frac{d\sigma^{^{\parallel,\bot(1)}}}{d\,Q^2d\,y} \ .
\end{equation}
The additional term on the right hand side of Eq.~(20) takes into
account 
RC due to real and virtual photon emission without its leading part that
is absorbed by $D$--functions. To find
$d\sigma^{^{\parallel,\bot(1)}}/d\,Q^2d\,y$, we must calculate the
corresponding cross--sections of the process (1) (with virtual and soft
corrections) and of the process (2), and then subtract from their sum
the right hand side of formula (3) with $$
\frac{d\sigma^{^{\parallel,\bot}}_{hard}}{d\,Q^2d\,y} =
\frac{d\sigma^{^{\parallel,\bot(B)}}}{d\,Q^2d\,y} \ , $$
which appears in the same order of the perturbation theory.

We begin with the calculation of the cross--section of the radiative
process (2) (the corresponding polarization calculations were performed
for the case of deep inelastic scattering \cite{DISpolarization})
\begin{equation}\label{21}
\frac{d\sigma^{^{\gamma(p)}}}{d\,Q^2d\,y}=\frac{2\pi\alpha^2(q^2)}{Vq^4}
\frac{\alpha}{4\pi^2}L^{^{\gamma}}_{\mu\nu}H_{\mu\nu}\frac{d^3k}{k_0}\frac{d^3p_2}
{p_{20}}\delta(p_1+k_1-k_2-p_2-k)\ ,
\end{equation}
where $q=k_1-k_2-k = p_2-p_1.$	In further we will  be interested in
the
polarization dependent parts of the leptonic $L_{\mu\nu}$ and hadronic
$H_{\mu\nu}$ tensors and assume that the degree of initial electron
polarization is
equal to $1$.
  In this
case we have
\begin{equation}\label{22,hadron tensor}
H_{\mu\nu} =
-iM\epsilon_{\mu\nu\lambda\rho}q_{\lambda}\bigl[-G_E(q^2)A_{\rho} +
\frac{2(G_E(q^2)-G_M(q^2))}{4M^2-q^2}(Ap_1)p_{1\rho}\bigr]G_M(q^2) \ ,
\end{equation}
\begin{equation}\label{23, lepton tensor}
L^{^{\gamma}}_{\mu\nu} =
-2i\epsilon_{\mu\nu\lambda\rho}q_{\lambda}[k_{1\rho}R_t +k_{2\rho}R_s] \ ,
\end{equation}
$$R_t= \frac{u+t}{st}-2m^2\bigl(\frac{1}{s^2}+\frac{1}{t^2}\bigr) \ , \
R_s= \frac{u+s}{st}-2m^2\frac{s_t}{ut^2}\ , s_t = \frac{-u(u+Vy)}{u+V}\ ,
$$
where $A$ is the 4--vector of the recoil proton polarization and we use
the following notation for invariants
$$u=(k_1-k_2)^2\ , \ \ s=2kk_2\ , \ \ t=-2kk_1\ , \ \ q^2=u+s+t\ , \ \
Q^2=-u\ . $$

It is convenient to express the recoil proton polarization 4--vector $A$ in
terms of the particle 4--momenta and Lorentz invariants. Below we use
the following parameterization for $A^{^{\parallel}}$ and $A^{^{\bot}}$
\begin{equation}\label{24, representation for polarization}
A^{\parallel}_{\mu} = \frac{2M^2q_{\mu}-q^2p_{2\mu}}{MQ_{\parallel}}\ , \
\ Q_{\parallel} = \sqrt{-q^2(4M^2-q^2)}\ ,
\end{equation}
\begin{equation}\label{25}
A^{\bot}_{\mu} =
\frac{2[2M^2k_1q+q^2k_1p_1]p_{2\mu}-2[2M^2k_1q-q^2k_1p_2]p_{1\mu}+q^2
(q^2-4M^2)k_{1\mu}}{2Q_{\bot}}\ ,
\end{equation}
$$Q_{\bot}=\sqrt{q^2[q^2M^2(k_1p_1+k_1p_2)^2+(2M^2k_1q-q^2k_1p_2)(2M^2k_1q
+q^2k_1p_1)]}\ .$$
$$2k_1p_2 = V +u+t\ , \ \ 2k_1q=u+t\ . $$

It is easy to verify that 4--vector $A^{^{\parallel}}$ in the rest frame
of the recoil proton has components $(0,\vec n),$ where 3--vector $\vec n$
has orientation of the recoil proton 3--momentum in laboratory system.
One can verify also that $A^{^{\bot}}A^{^{\parallel}}=0$ and in the rest
frame of the recoil proton
$$A^{^{\bot}} = (0,\vec n_{\bot})\ , \ \ \vec n_{\bot}^2 = 1\ ,  \ \
\vec n \vec n_{\bot} =0 \ , $$
where the 3--vector $\vec n_{\bot}$ is within the plane $(\vec
k_1,
\vec p_2)$ in the laboratory system.

For the case of longitudinal polarization, the contraction of
leptonic and hadronic tensors yields \begin{equation}\label{26,
contraction, long}
\frac{L_{\mu\nu}^{\gamma}H_{\mu\nu}}{q^4}=-\frac{2m^2}{s^2}(q_s^2+2V)F(q_s^2)
-\frac{2m^2}{t^2}(u+2V)\bigl(1+\frac{s_tq_t^2}{u^2}\bigr)F(q_t^2)+
\end{equation}
$$\bigl\{[\frac{1}{tu}[(u^2+q^4)(u+2V)-2q^2(q^2-q_t^2)(u+V)] +
\frac{1}{sq_s^2}[(q^4+u^2)(q_s^2+2V)
-2q^2V(q^2-q_s^2)]\bigr\}\frac{F(q^2)}{q^2-u} \ ,$$
where
$$q_t^2 = u+s_t =\frac{uV(1-y)}{u+V}\ , \ \ q_s^2 = u+t_s =
\frac{uV}{V(1-y)-u}\ , \ \ t_s = \frac{u(u+Vy)}{V(1-y)-u} \ , $$
$$F(q^2) = -G_M^2(q^2)\frac{1}{q^2}\sqrt{\frac{-q^2}{4M^2-q^2}}\ .$$
The physical meaning of quantities $q_t^2$ and $q_s^2$ is as follows:
$q_t^2$ and $q_s^2$ are the values of $q^2$ in the cases of the
initial--state and final--state collinear
radiation, respectively. When writing the formula (26), we took into
account the fact that
terms containing the electron mass squared contribute only in collinear
kinematics.

To separate the contribution into the right-hand side of Eq.~(26) due to
collinear radiation for the pole--like terms, we apply the operations
$\hat P_t$ and $\hat P_s$, $$\frac{1}{t}f(q^2,u,t,s) =\frac{1}{t}(1-\hat P_t +\hat
P_t)f(q^2,u,t,s)\ , \ \ \hat P_tf(q^2,u,s,t) = f(q_t^2,u,s_t,0)$$
for arbitrary nonsingular function at $t\rightarrow 0$ and similarly
for
$1/s$ terms. Therefore, we can rewrite the right hand side of
Eq.~(26) in the form
\begin{equation}\label{27}
\Bigl\{-\frac{2m^2}{s^2}(q_s^2+2V)\hat P_s
-\frac{2m^2}{t^2}(u+2V)\bigl(1+\frac{s_tq_t^2}{u}\bigl)\hat
P_t\Bigr\}F(q^2) +\Bigl\{\frac{(u+2V)(u^2+q_t^4)}{ut}\hat P_t +
\end{equation} $$\frac{(q_s^2+2V)(u^2+q_s^4)}{q_s^2s}\hat P_s +
\frac{1-\hat P_t}{ut}[(u+2V)(u^2+q^4)-2q^2(q^2-q_t^2)(u+V)] + $$
$$\frac{1-\hat P_s}{q_s^2s}[(q_s^2+2V)(u^2+q^4)-2q^2V(q^2-q_s^2)]\Bigr\}
\frac{F(q^2)}{q^2-u} \ . $$

For the case of transverse polarization the contraction of leptonic and
hadronic tensors is more complicated,
\begin{equation}\label{perpendicular polarization,28}
\frac{1}{q^4}L_{\mu\nu}^{\gamma}H_{\mu\nu} =
\bigr\{[q^2(u+t+2V)^2+(4M^2-q^2)(u+t)^2]R_t + [q^2(u+t+2V)(t-q^2+2V( 1-y))+
\end{equation}
$$(4M^2-q^2)(uq^2-st)]R_s\bigr\}
\frac{G_E(q^2)G_M(q^2)}{q^4}
\sqrt{\frac{-q^2M^2}{(4M^2-q^2)(-q^2V(V+u+t)-M^2(u+t)^2)}} \ .$$
The expression in the round brackets on the right hand side of Eq.~(28)
can
be rewritten in the form suitable for the photon angular integration as
follows:
\begin{equation}\label{29}
-2[q^2Vy+4M^2(q^2+u)]-\frac{2m^2}{s^2}4V^2q_s^2K_s\hat P_s
-\frac{2m^2}{t^2}4V^2q_t^2\bigl(1+\frac{s_tq_t^2}{u^2}\bigr)K_t\hat P_t+
\end{equation}
$$\frac{1}{t}\bigl[\frac{4V^2q_t^2(u^2+q_t^2)}{u(q_t^2-u)}K_t\hat P_t+(1-
\hat
P_t)\frac{q^2}{u(q^2-u)}[4V^2(u^2+q^4)K_q-2q^2(q^2-q_t^2)(u+2V)(u+V)]
\bigr]+$$
$$\frac{1}{s}\bigl[\frac{4V^2(u^2+q_s^2)}{(q_s^2-u)}K_s\hat P_s+(1- \hat
P_s)\frac{q^2V}{q_s^2(q^2-u)}[4V(u^2+q^4)K_s-2V(q^2-q_s^2)(2Vq^2-u^2)]
\bigl]\ , $$
$$K_s=1+\frac{q_s^2}{V}(1+\tau)\ , \ \ K_t =1+\frac{u}{V}+\frac{u^2\tau}
{Vq_t^2}\ , \ \ K_q=1+\frac{u}{V}+\frac{u^2\tau}{Vq^2}\ . $$

To perform the photon angular integration we choose the system $\vec
k_1+\vec p_1-\vec k_2 =0.$ In this system the energies of particles are
\begin{equation}\label{30,energies}
k_0=\frac{a}{2\sqrt{R}}\ , \ k_{10}=\frac{u+V}{2\sqrt{R}}\ , \
k_{20}=\frac{V(1-y)-u}{2\sqrt{R}}\ , \ p_{10}=\frac{2M^2+Vy}{2\sqrt{R}}\ ,
\ p_{20}=\frac{R+M^2}{2\sqrt{R}}\ ,
\end{equation}
$$a= u+Vy\ , \ \ R =a+M^2\ .$$
Taking the OZ axis along the initial proton 3--momentum in the chosen
system we also have
\begin{equation}\label{31,angles}
c_k=\cos{\theta_k} = \frac{2M^2-2p_{10}p_{20}-q^2}{2|\vec p_1||\vec p_2|}\
, \
c_2=\cos{\theta_2} = \frac{2k_{20}p_{10}-V(1-y)}{2|\vec p_1||\vec k_2|} \ ,
\end{equation}
$$c_1=\cos{\theta_1} = \frac{2k_{10}p_{10}-V}{2|\vec p_1||\vec k_1|} \ ,
  |\vec p_1| = \frac{\sqrt{V^2y^2-4uM^2}}{2\sqrt{R}}\ , \ |\vec p_2|=
  k_0 \ , $$
where $\theta_1(\theta_2)$ is the polar angle of the initial (scattered)
electron and $\theta_k$ is the photon polar angle. Besides Eqs. (30) and
(31) we will use the relation \begin{equation}\label{32}
\frac{d^3k}{k_0}\frac{d^3p_2}{p_{20}}\delta(k_1+p_1-k_2-k-p_2) =
\frac{a}{2R}d\varphi\,d\cos{\theta_k}\ .
\end{equation}

Let us concentrate on the case with longitudinal polarization of the
recoil proton. For the terms containing $m^2/s^2,$ $m^2/t^2,$ $\hat P_t/
t$ and $\hat P_s/s$ we can use the following formulae
\begin{equation}\label{33,formulae for integration}
\int\frac{m^2d\varphi d\cos{\theta_k}}{2\pi s^2} =\int\frac{m^2d\varphi
d\cos{\theta_k}}{2\pi t^2}=\frac{2R}{a^2} \ , \
\int\frac{d\varphi d\cos{\theta_k}}{2\pi s}=
\frac{2R}{a(V(1-y)-u)}(L_s+L)\ ,
\end{equation}
$$\int\frac{d\varphi d\cos{\theta_k}}{2\pi (-t)}=
\frac{2R}{a(u+V)}(L_t+L)\ , \ L_s = \ln\frac{(V(1-y)-u)^2}{-uR} \ , \
L_t = \ln\frac{(V+u)^2}{-uR} \ . $$

Terms which contain $(1-\hat{P_t})\ , \ (1-\hat{P_s})$ operators  can
be integrated over the azimuthal angle
and keep the integration with respect to $q^2$
 using
$d\,\cos{\theta_k} = {d\,q^2}/{2|\vec p_1||\vec p_2|}\
$,
\begin{equation}\label{34} \int\frac{d\,\varphi}{2\pi s 2|\vec p_1||\vec
p_2|}= \frac{2R}{a|q^2-q_s^2|(V(1-y)-u)}\ , \ \int\frac{d\,\varphi}{2\pi
(-t) 2|\vec p_1||\vec p_2|}= \frac{2R}{a|q^2-q_t^2|(V+u)}\ .
\end{equation}
The limits of $q^2$--integration in this case can be derived from
the restriction on $\cos{\theta_k}$ in the chosen system;
$|\cos{\theta_k}|<1\ .$ This restriction leads to the relation
\begin{equation}\label{35,limits for q^2}
q_-^2<q^2<q_+^2 \ , \ \ q_{\pm}^2=\frac{1}{2R}\bigl[2uM^2-Vy(u+Vy)\pm
(u+Vy)\sqrt{V^2y^2-4uM^2}\bigr] \ .
\end{equation}

By using Eqs. (33), (34) and (35) we can write the cross--section of
the
radiative process (2) in the case of longitudinal polarization of the
recoil proton as follows
\begin{equation}\label{36}
\frac{d\sigma^{^{\parallel\gamma}}}{d\,Q^2d\,y}=\frac{2\alpha}{V}
\Bigl\{-\frac{q_s^2+2V}{u+Vy}\hat P_s
-\frac{(u+2V)(u^2+s_tq_t^2)}{u^2(u+Vy)}\hat P_t -
\end{equation}
$$-[1+L_t+(L-1)]\frac{(u+2V)(u^2+q_t^4)}{2u(u+V)(q_t^2-u)}\hat P_t+
[1+L_s+(L-1)]\frac{(q_s^2+2V)(u^2+q_s^4)}{2q_s^2(V(1-y)-u)(q_s^2-u)}\hat
P_s+$$
$$\int\limits_{q_-^2}^{q_+^2}\Bigl[
-\frac{d\,q^2}{|q^2-q_t^2|}(1-\hat P_t)
\frac{(u+2V)(u^2+q^4)-2q^2(q^2-q_t^2)(u+V)}{2u(u+V)(q^2-u)}+ $$
$$\frac{d\,q^2}{|q^2-q_s^2|}(1-\hat P_s)
\frac{(q_s^2+2V)(u^2+q^4)-2q^2V(q^2-q_s^2)}{2q_s^2(V(1-y)-u)(q^2-u)}
\Bigr]\Bigr\}
\alpha^2(q^2)F(q^2)\theta\bigl(y+\frac{u}{V}-\frac{2M\Delta\varepsilon}{V}
\bigr)\ .  $$

The appearance of the $\theta$--function on the right side of Eq.~(36)
is connected with the restriction on the photon hardness in the
radiative process (2)
\begin{equation}\label{37}
k_0 = \frac{u+Vy}{2\sqrt{M^2+u+Vy}}>\Delta\varepsilon \ \ \rightarrow
 \ \ y>-\frac{u}{V}+\frac{2M\Delta\varepsilon}{V}\ ,
\end{equation}
where $\Delta\varepsilon$ is the the minimal photon energy in the chosen
coordinate system.

To be complete, we should also take into account the RC due to virtual
and
soft (with the energy smaller than $\Delta\varepsilon$) photon emission to
the cross--section of the elastic process (1). It can be written as (see,
for example, \cite{KurMerFad}) \begin{equation}\label{38}
\frac{d\sigma^{^{\parallel(V+S)}}}{d\,Q^2d\,y}=\frac{4\pi\alpha^2(-Q^2)}
{V}\bigl(1-\frac{Q^2}{2V}\bigr)F(-Q^2)\frac{\alpha}{2\pi}\Bigl[2(L-1)\bigl(
\ln\frac{4M^2(\Delta\varepsilon)^2}{V(u+V)}+\frac{3}{2}\bigr)-
\end{equation}
$$-1-\frac{\pi^2}{3}-\ln^2\frac{u+V}{V}-2f\bigl(\frac{u+V+u\tau}{u+V}\bigr)
\Bigr]\delta\bigl(y-\frac{Q^2}{V}\bigr) \ , \ f(x)=\int\limits_0^x
\frac{d\,x}{x}\ln(1-x). $$

Therefore, the sum of the cross--sections of the processes (1) and (2) is
defined by the formula
\begin{equation}\label{39,sum}
\frac{d\sigma^{^{\parallel(B)}}}{d\,Q^2d\,y}+
\frac{d\sigma^{^{\parallel\gamma}}}{d\,Q^2d\,y}+
\frac{d\sigma^{^{\parallel(S+V)}}}{d\,Q^2d\,y} \ .
\end{equation}

To include the hard cross--section into the electron structure
function
representation (3) in the form (39) and get rid of the double counting,
we must remove from the sum (39) the contribution which arises in the
representation (3) in the first order with respect to fine
structure constant $\alpha$ at
$$\frac{d\sigma^{^{\parallel}}_{hard}}{d\,Q^2d\,y} =
\frac{d\sigma^{^{\parallel(B)}}}{d\,Q^2d\,y}\ . $$
The procedure for finding this contribution is described in
\cite{KurMerFad}. We can verify that it equals to
\begin{equation}\label{40, doubling term}
\frac{2\alpha}{V}\Bigl\{(L-1)\Bigl[-\frac{(u+2V)(u^2+q_t^4)}{2u(u+V)(q_t^2-u)}
\hat P_t+\frac{(q_s^2+2V)(u^2+q_s^4)}{2q_s^2(V(1-y)-u)(q_s^2-u)}\hat P_s\Bigr]
\alpha^2(q^2)F(q^2)
\end{equation}
$$\times
\theta\bigl(y+\frac{u}{V}-\frac{2M\Delta\varepsilon}{V}
\bigr)+2(L-1)\bigl(\ln\frac{4M^2(\Delta\varepsilon)^2}{V(u+V)}+
\frac{3}{2}\bigr)\bigl(1-\frac{Q^2}{2V}\bigr)\alpha^2(-Q^2)F(-Q^2)\delta\bigl
(y+\frac{u}{V}\bigr) \Bigr\} \ . $$

Thus, we can write the final result for the
$d\sigma^{^{\parallel}}_{hard}/d\,Q^2d\,y$ in the following very compact
form
\begin{equation}\label{41, final ||}
\frac{d\sigma^{^{\parallel}}_{hard}}{d\,Q^2d\,y}=\frac
{d\sigma^{^{\parallel(B)}}}{d\,Q^2d\,y}\bigl\{1+\frac{\alpha}{2\pi}\bigl[
-1-\frac{\pi^2}{3}-\ln^2\frac{u+V}{V}-2f\bigl(\frac{u+V+u\tau}{u+V}\bigr)
\bigr]\bigr\}+
\end{equation}
$$\frac{2\alpha}{V}\Bigl\{\frac{(u+2V)(q_t^2-u)}{2u(u+V)}\hat P_t +
\frac{(q_s^2+2V)(q_s^2-u)}{2uV}\hat P_s +P\int\limits_{q_-^2}^{q_+^2}
\frac{d\,q^2}{q^2-u}\Bigl[\frac{1}{|q^2-q_s^2|}(1-\hat P_s)$$
$$\times\frac{(q_s^2+2V)(u^2+q^4)-2q^2V(q^2-q_s^2)}{2q_s^2(V(1-y)-u)}-$$
$$\frac{1}{|q^2-q_t^2|}(1-\hat P_t)
\frac{(u+2V)(u^2+q^4)-2q^2(u+V)(q^2-q_t^2)}{2u(V+u)}\Bigr]\Bigr\}
\alpha^2(q^2)F(q^2)\theta\bigl(y+\frac{u}{V}\bigr) \ , $$
where $P$ is the symbol of the principal value integration. When writing
the last formula, we used the following relations
\begin{equation}\label{42}
P\int\limits_{q_-^2}^{q_+^2}\frac{d\,q^2[f(q^2)-f(q_t^2)]}{|q^2-q_t^2|
(q^2-u)} = \frac{f(q_t^2)}{q_t^2-u}L_t +\int\limits_{q_-^2}^{q_+^2}\frac
{d\,q^2}{|q^2-q_t^2|}\Bigl(\frac{f(q^2)}{q^2-u}-\frac{f(q_t^2)}{q_t^2-u}
\Bigr) \ ,
\end{equation}
\begin{equation}\label{43}
P\int\limits_{q_-^2}^{q_+^2}\frac{d\,q^2[f(q^2)-f(q_s^2)]}{|q^2-q_s^2|
(q^2-u)} = \frac{f(q_s^2)}{q_s^2-u}L_s +\int\limits_{q_-^2}^{q_+^2}\frac
{d\,q^2}{|q^2-q_s^2|}\Bigl(\frac{f(q^2)}{q^2-u}-\frac{f(q_s^2)}{q_s^2-u}
\Bigr) \ ,
\end{equation}
where the symbol $P$ indicates how one shall integrate the unphysical
singularity at $q^2=u.$ These relations allow to see that infrared
singularities of separate terms
in $d\sigma^{^{\parallel(1)}}/dQ^2dy$ exactly cancel each other.  That
is why
we omitted from argument of the $\theta$--function on the right side of
Eq.~(41) the term $-2M\Delta\varepsilon/V.$ For numerical
calculations the symbol $P$ can be understood as
$$
P\int\limits_{q_-^2}^{q_+^2}\frac{d\,q^2}{q^2-u}F(q^2)=
\int\limits_{q_-^2}^{q_+^2}\frac{d\,q^2}{q^2-u}\bigl(F(q^2)-F(u)\bigr)
+F(u)\log{q_+^2-u \over q_-^2-u}
$$

The hard part of the cross--section in the case of transverse
polarization of the recoil proton can be derived in full analogy with the
above. The main difference is caused by the fact that the vector of
transverse polarization has complicated dependence on the
photon azimuthal
angle $\phi$ and therefore  even $\phi$ integration becomes
nontrivial. The straightforward calculations give
\begin{equation}\label{44}
\frac{d\sigma^{^{\bot}}_{hard}}{d\,Q^2d\,y}=\frac
{d\sigma^{^{\bot(B)}}}{d\,Q^2d\,y}\bigl\{1+\frac{\alpha}{2\pi}\bigl[
-1-\frac{\pi^2}{3}-\ln^2\frac{u+V}{V}-2f\bigl(\frac{u+V+u\tau}{u+V}\bigr)
\bigr]\bigr\}+
\end{equation}
$$\frac{2\alpha}{V}\Bigl\{\Bigl[\frac{2(q_t^2-u)V}{u(u+V)}+\frac{2V(u^2+q_t^4)}
{u^2(u+Vy)}L_t\Bigr]\frac{K_t\hat P_t}{q_t^2}+
\Bigl[\frac{2(q_s^2-u)}{u}+\frac{2V(u^2+q_s^4)}{uq_s^2(u+Vy)}L_s\Bigr]
\frac{K_s\hat P_s}{q_s^2} + $$
$$\int\limits_{q_-^2}^{q_+^2}\frac{d\,q^2}{\sqrt{V^2y^2-4uM^2}}
\int\limits_0^{2\pi}\frac{d\varphi}{2\pi}
\Bigl[\frac{[-yq^2-4\tau(u+q^2)]}{q^4}+\frac
{1-\hat P_t}{t}
\Bigl(\frac{2V(u^2+q^4)}{uq^2(q^2-u)}K_q-$$
$$\frac{(u+2V)(u+V)(q^2-q_t^2)}{uV(q^2-u)}\Bigr) +
\frac{1-\hat P_s}{s}\Bigl(\frac{2V(u^2+q^4)}{q_s^2q^2(q^2-u)}K_s+
\frac{(u^2-2q^2V)(q^2-q_s^2)}{q_s^2q^2(q^2-u)}\Bigr)\Bigr]\Bigr\}$$
$$\times\alpha^2(q^2)\sqrt{\frac{M^2}{4M^2-q^2}}\bigl(1+\frac{u+t}{V}+
\frac{(u+t)^2\tau}{Vq^2}\bigr)^{-1/2}G_E(q^2)G_M(q^2)\theta\bigl(y+
\frac{u}{V}\bigr)\ . $$
For invariants $s$ and $t$ on the right side of Eq.~(44), we can neglect
the electron mass and use here the simplified expressions
\begin{equation}
s=c_{2i}-s_i\cos{\varphi}, \
-t=c_{1i}-s_i\cos{\varphi}, \
\end{equation}
$$
c_{1i}= 2k_0k_{10}[1-\cos{\theta_1}\cos{\theta_k}],\
c_{2i}= 2k_0k_{20}[1-\cos{\theta_2}\cos{\theta_k}],\
$$
$$
s_i= 2k_0k_{10}\sin{\theta_1}\sin{\theta_k}=
 2k_0k_{20}\sin{\theta_2}\sin{\theta_k}
$$

The integrals over $\phi$ can be performed in terms of elliptic
functions ${\cal K}$ and $\Pi$
\begin{eqnarray}\label{iii}
&&\int\limits_0^{2\pi}\frac{d\varphi}{2\pi}
\bigl(1+\frac{u+t}{V}+\frac{(u+t)^2\tau}{Vq^2}\bigr)^{-1/2}
=J_0=
\frac{2}{\pi\sqrt{X}}{\cal K}(\kappa)
\\
&&\int\limits_0^{2\pi}\frac{d\varphi}{2\pi t}
\bigl(1+\frac{u+t}{V}+\frac{(u+t)^2\tau}{Vq^2}\bigr)^{-1/2}
={J_t\over |q^2-q^2_t|}=\nonumber 
\\ && \qquad\qquad =
-\frac{B_t(1+b_t)\sqrt{\lambda\bar y}}{2(V+u)|q^2-q^2_t|
\sqrt{X}}\Biggl(\frac{2}{\pi}\sqrt{1-b_{1t}}{\cal
K}(\kappa)+\frac{B_t}{b_{1t}\bar y}\frac{1-\Lambda(\epsilon,\kappa)}
{\sqrt{1-\kappa^2/b_{1t}}
}\Biggr),
\nonumber\\
&&\int\limits_0^{2\pi}\frac{d\varphi}{2\pi s}
\bigl(1+\frac{u+t}{V}+\frac{(u+t)^2\tau}{Vq^2}\bigr)^{-1/2}
={J_s\over |q^2-q^2_s|}=\nonumber 
\\ && \qquad\qquad =
\frac{B_s(1+b_s)\sqrt{\lambda\bar y}}{2(V-a)|q^2-q^2_s|
\sqrt{X}}\Biggl(\frac{2}{\pi}\sqrt{1-b_{1s}}{\cal
K}(\kappa)+\frac{B_s}{b_{1s}\bar y}\frac{1-\Lambda(\epsilon,\kappa)}
{\sqrt{1-\kappa^2/b_{1s}}
}\Biggr),
\nonumber
\end{eqnarray}
where
$$
X=(1+x_+)(1-x_-){M^2s_i^2\over -q^2V^2}, \ \kappa^2={2(x_+-x_-)\over
(1+x_+)(1-x_-)},
\ \bar y={-x_-+1\over -x_--1},
$$

\begin{eqnarray}
2M^2s_i x_\pm &=& 2M^2(-q^2+c_{2i})-Vq^2(1\pm\sqrt{1-4M^2/q^2})
\nonumber\\
&=& 2M^2(-u+c_{1i})-Vq^2(1\pm\sqrt{1-4M^2/q^2})
\end{eqnarray}

$$
b_{1s,t}={1+\bar y-b_{s,t}+b_{s,t}\bar y\over \bar y(1+b_{s,t})},\
b_{s,t}={c_{2,1i}\over s_i}, \ B_{s,t}=b_{1s,t}\bar y+1-\bar y
.$$

The function
$\Lambda(\epsilon,\kappa)$ ($\epsilon=\arcsin((1-b_1)/(1-\kappa^2))$)
is non-singular Heuman's Lambda function varying from 0 to 1
(see \cite{abste} for
details and exact definitions). It is related with complete elliptic integral
$\Pi(b_1,\kappa)$ of the third kind
\begin{eqnarray}\label{Lambda}
\frac{2}{\pi}\Pi(b_1,\kappa)&=&
	     \frac{1-\Lambda(\epsilon,\kappa)}
	     {\sqrt{1-b_1}\sqrt{1-\kappa^2/b_1}}
	    +\frac{2}{\pi}{\cal K}(\kappa)
\end{eqnarray}
For $\epsilon \rightarrow 0$ (or
$b_1 \rightarrow 1$) this function goes to zero.
In the last formula singular behavior of $\Pi(b_1,\kappa)$ for
$b_1\rightarrow 1$ is extracted explicitly in the first term.
This limit
corresponds to collinear radiation:
\begin{eqnarray}
\sqrt{1-b_{1t}}&=&{u+V\over (1+b_t)s_i \sqrt{\bar y\lambda}}|q^2-q^2_t|
\nonumber \\
\sqrt{1-b_{1s}}&=&{V-a\over (1+b_s)s_i \sqrt{\bar y\lambda}}|q^2-q^2_s|
\end{eqnarray}
where $\lambda=y^2V^2-4M^2u$.
As a result of substituting Egs (\ref{iii}--\ref{Lambda}) into the
formula for hard cross--section (\ref{44})  we arrive at the
same structure of singularities
as in the longitudinal case (\ref{36}). 
 In the collinear limit $q^2\rightarrow q^2_{t,s}$, we have
$$
\bar y X (1-\kappa^2/b_{1t,s}) \rightarrow K_{t,s} \qquad
b_{t,s},b_{1t,s},B_{t,s} \rightarrow 1, \qquad
\Lambda(\epsilon,\kappa) \rightarrow 0
$$
These limiting formulae allows us to use relations 
 (\ref{42},\ref{43}) 
to
write the final expression for hard cross section in such a form that
provides an explicit cancellation of infrared divergence
 in the same way as in the case of the
longitudinal polarization.

Combining all results
together, we obtain the
final formula for cross section in the transversely polarized case:

\begin{equation}\label{final2}
\frac{d\sigma^{^{\bot}}_{hard}}{d\,Q^2d\,y}=\frac
{d\sigma^{^{\bot(B)}}}{d\,Q^2d\,y}\bigl\{1+\frac{\alpha}{2\pi}\bigl[
-1-\frac{\pi^2}{3}-\ln^2\frac{u+V}{V}-2f\bigl(\frac{u+V+u\tau}{u+V}\bigr)
\bigr]\bigr\}+
\end{equation}
$$\frac{2\alpha}{V}\Bigl\{	\frac{2(q_t^2-u)V}{u(u+V)}
		    \frac{K_t\hat P_t}{q_t^2}+
      \frac{2(q_s^2-u)}{u}
\frac{K_s\hat P_s}{q_s^2} +
P\int\limits_{q_-^2}^{q_+^2}\frac{d\,q^2}{\sqrt{\lambda}(q^2-u)}
\Bigl[\frac{[-yq^2-4\tau(u+q^2)]}{q^4}(q^2-u)J_0
$$
$$
+\frac{1-\hat P_t}{|q^2-q^2_t|}J_t
\Bigl(\frac{2V(u^2+q^4)}{uq^2	    }K_q-
\frac{(u+2V)(u+V)(q^2-q_t^2)}{uV       }\Bigr) +
$$
$$
+\frac{1-\hat P_s}{|q^2-q^2_s|}J_s
\Bigl(\frac{2V(u^2+q^4)}{q_s^2q^2	}K_s+
\frac{(u^2-2q^2V)(q^2-q_s^2)}{q_s^2q^2	     }\Bigr)\Bigr]\Bigr\}$$
$$\times\alpha^2(q^2)\sqrt{\frac{M^2}{4M^2-q^2}}
G_E(q^2)G_M(q^2)\theta\bigl(y+
\frac{u}{V}\bigr)\ . $$


The theoretical formula for the ratio
of longitudinal and transverse polarizations of the
recoil proton that was measured in recent experiments
\cite{Bates,1CEBAFothers} is defined by the ratio of the right--hand side of
Eq.~(3) for longitudinal polarization (with (41) as the hard cross-section
under integral sign) and for transverse one (with (44) as the hard
cross--section). This high precision formula takes into account model
independent RC with all the leading and the main part of the
next--to--leading corrections, and has accuracy at the level of per
mile.

\section{Numerical analysis}

The  ratio of proton elastic
formfactors $G_{ep}/G_{mp}$ measured experimentally  
\cite{Bates,1CEBAFothers} is related
to the ratio
of recoiled proton polarization components. At the Born level 
 (i.e. without RC) the
ratio of
polarizations is defined by the ratio of spin dependent cross section given
by
(\ref{16}) and (\ref{17}):
\begin{equation}\label{ratiodef}
\frac{P_T}{P_L}=\frac{\sigma_T^0}{\sigma_L^0}
\end{equation}

\begin{figure}[!t]
\unitlength 1mm
\begin{center}
\begin{picture}(160,80)
\put(0,-10){
\epsfxsize=7cm
\epsfysize=7cm
\epsfbox{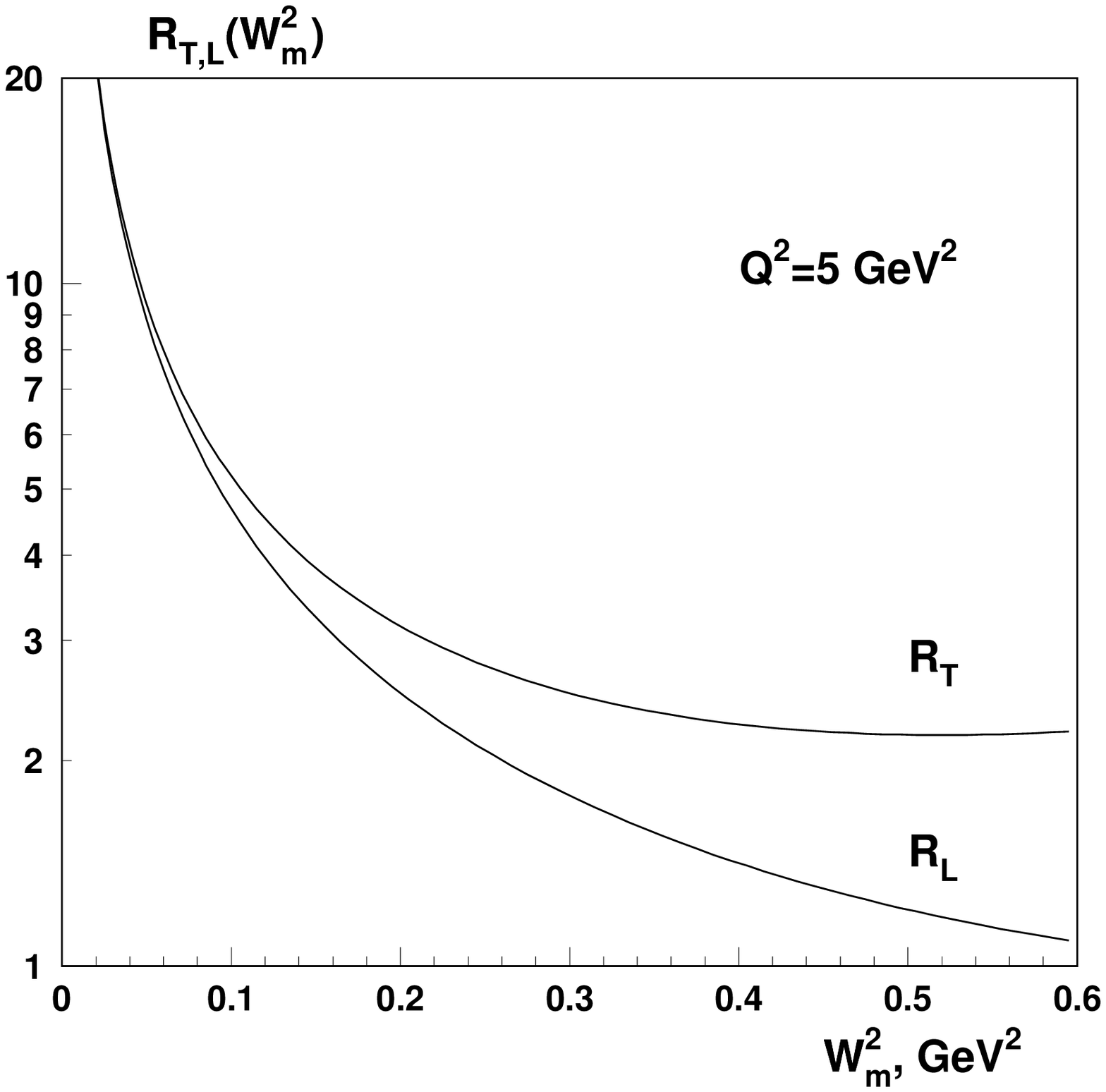}
}
\put(80,-10){
\epsfxsize=7cm
\epsfysize=7cm
\epsfbox{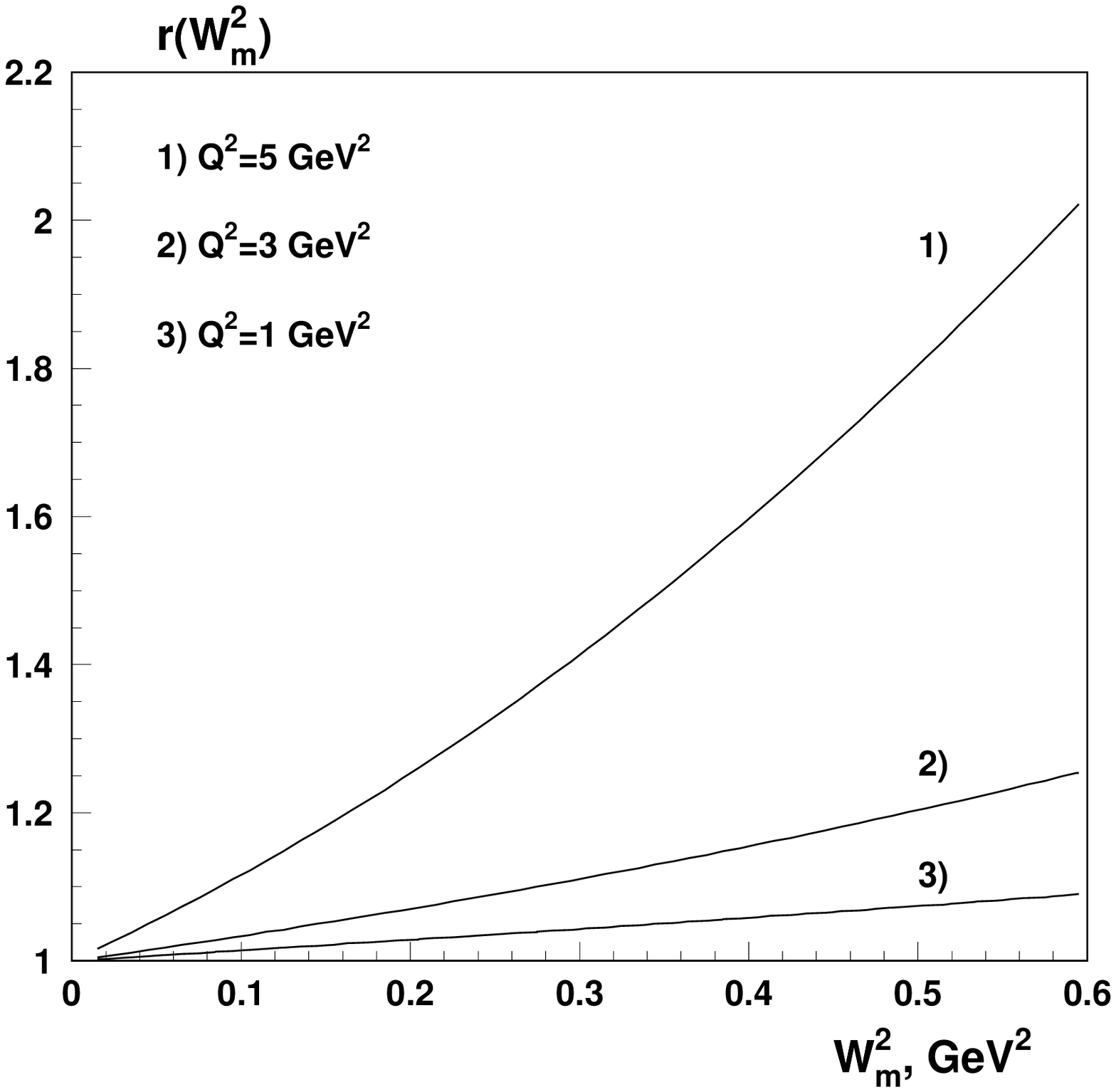}
}
\end{picture}
\end{center}
\caption{\label{sigtl} Longitudinal and transverse polarization
parts of cross sections normalized to born ones (left plot) and their ratios
(right plot) (see Eq.({\ref{eqsi}})
for exact definitions) as a function of missing mass squared for
beam energy 4.26 GeV ($V$=8GeV$^2$).
}
\end{figure}

The photon spectrum can be defined as a function of missing mass
$W^2_m=yV-Q^2$
(either $y$ or photon
energy in the  chosen
frame $E_\gamma$) of observed cross section  $\sigma_{T,L}(W_M^2)$ defined 
by master
equation (\ref{3}).
 An integral over
$y$ gives a radiative correction to recoil polarizations and to
their ratio. Let us define the following quantities
\begin{equation}\label{eqsi}
R_{T,L}(W_m^2)=\frac{\sigma_{T,L}(W_m^2)}{\sigma_{T,L}^0},
\quad
r(W_m^2)=\frac{R_{T}(W_m^2)}{R_{L}(W_m^2)},
\quad
r_{T,L}=\int \frac{dW^2_m}{V} R_{T,L}(W_m^2),
\quad
r=\frac{r_{T}}{r_{L}}.
\end{equation}

In Figure \ref{sigtl} the $R_{T,L}$ as a function of missing mass is presented.
For very small values of missing mass or alternatively	for $y\rightarrow Q^2/V$
the cross sections reproduce the $\delta$-function behavior. In the limit
(\ref{18}) there are three delta-functions (from $D^u$, $D^p$ and from
$y$-dependence of Born
cross section) and only 
two integration. So we have behavior as in Eq. (\ref{18})  in
this limit. Only the 
factorization part is important here, so both longitudinal and
transverse $R's$ are practically the same. For larger values of $W^2_m$ (or $y$)
nonfactorized part contribution becomes important. It can be seen from Figure
\ref{sigtl}b, where ratios of these spectrum are presented.

\begin{figure}[!t]
\unitlength 1mm
\begin{center}
\begin{picture}(160,80)
\put(0,-10){
\epsfxsize=7cm
\epsfysize=7cm
\epsfbox{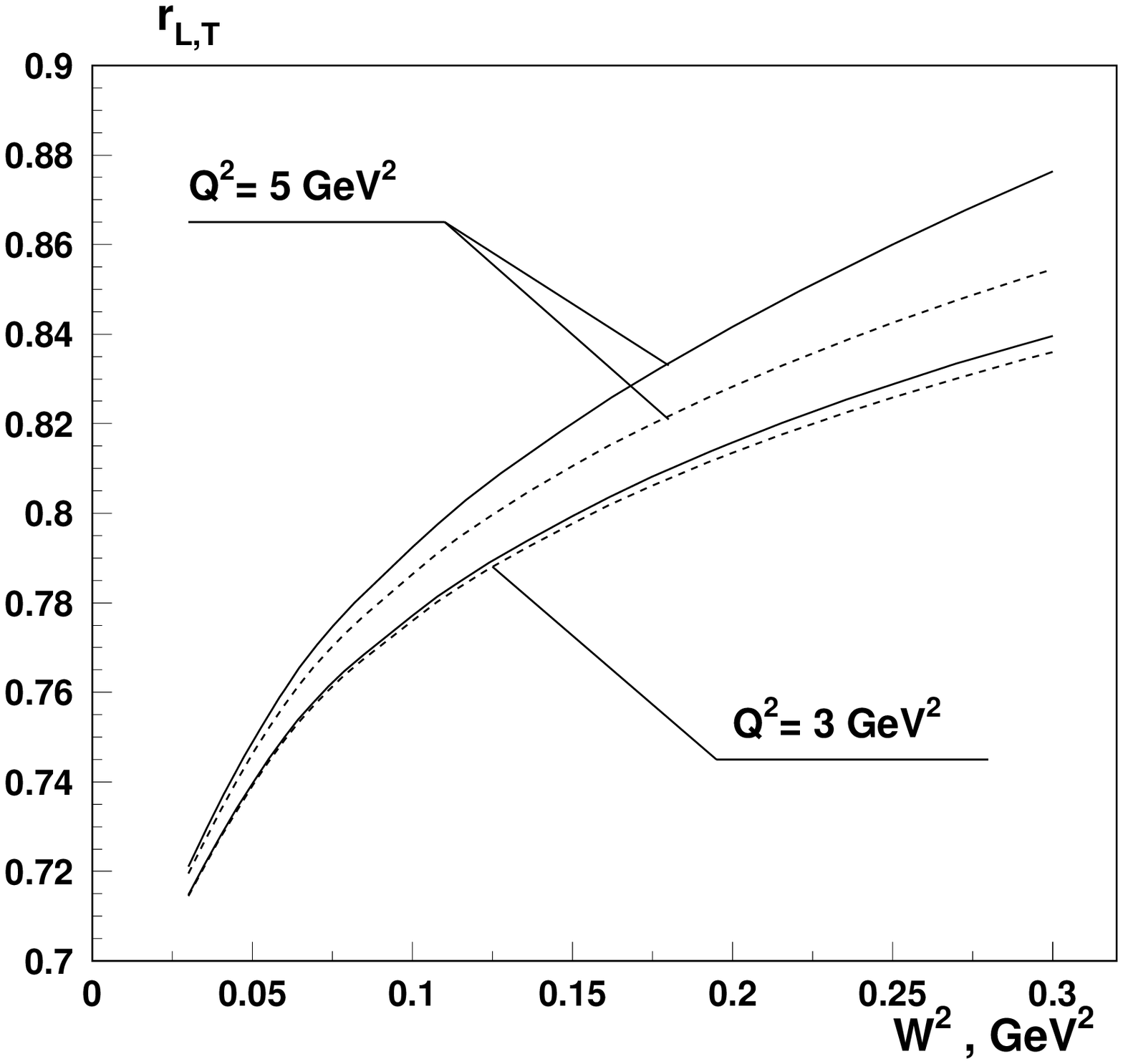}
}
\put(80,-10){
\epsfxsize=7cm
\epsfysize=7cm
\epsfbox{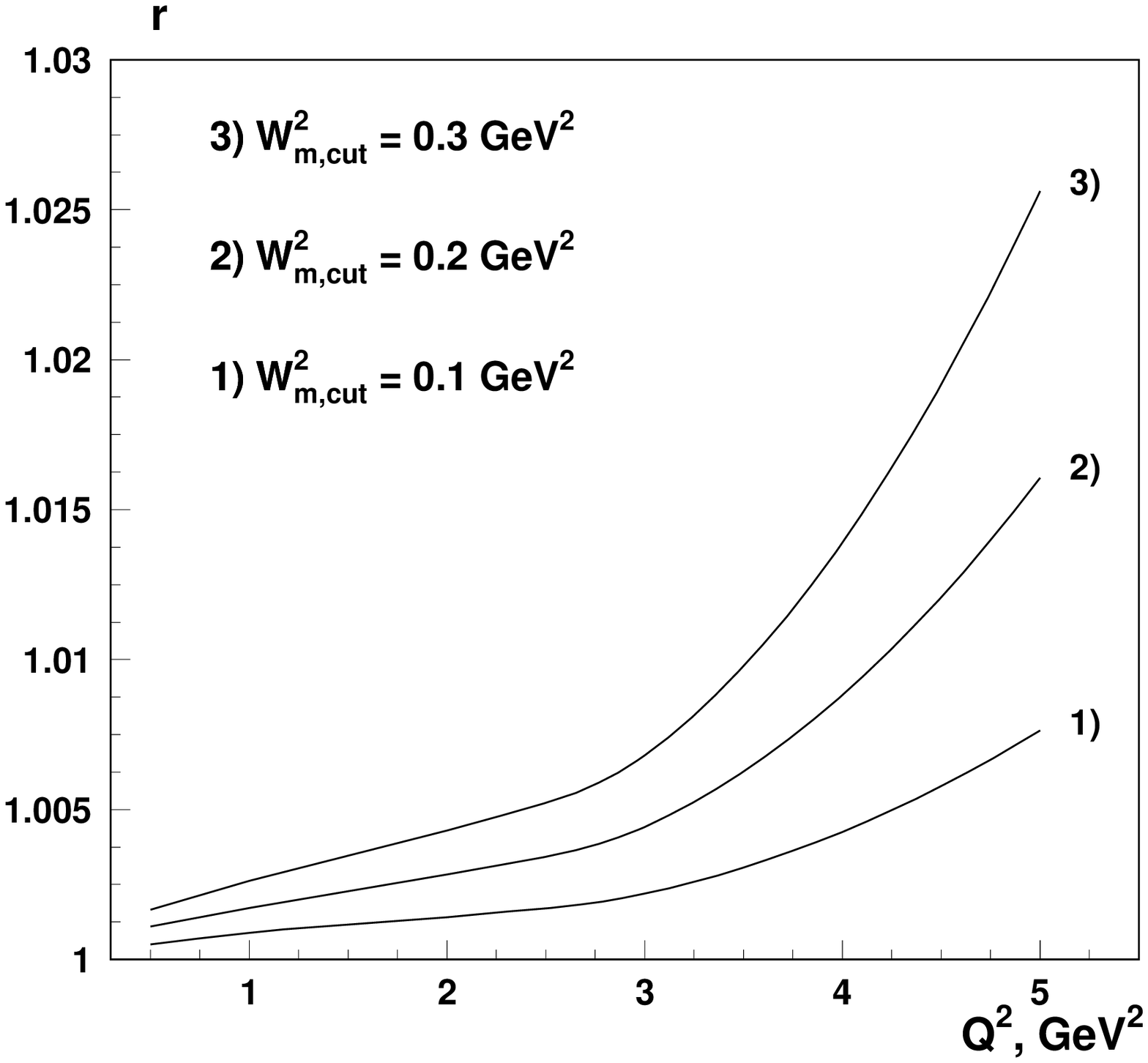}
}
\end{picture}
\end{center}
\caption{\label{rec3}
Radiative correction to recoil polarization rations $r_{T,L}$ (left plot) and
 $r$ (right plot) (\ref{eqsi})
 within the kinematical conditions of JLAB, as a function of $Q^2$
 and value of
 a cut on missing mass for
beam energy 4.26 GeV ($V$=8GeV$^2$).
 Solid (dashed) line on the left plot shows $r_T$
 ($r_L$).
}
\end{figure}

Figure \ref{rec3} presents the results integrated over $dy=dW^2_m/V$. This
integration has to be performed up to some specific values of a cut on the 
missing
mass which is defined by experimental conditions. Using the hard cut leads to
negative values RC (or $r_{T,L}$ becomes less then one), because 
the contribution of loops, which is usually negative, dominates
in this case. If the positive contribution of hard photon radiation is allowed by
using less stringent cuts, the radiative correction to polarized parts of cross
section goes up and can exceed several tens of per cents. The right plot in
 Figure
\ref{rec3} gives a radiative correction factor to the polarization ratio or the 
measured
ratio of formfactors. One can see that the radiative correction to it is
rising
not only with the increasing  value of the cut but also 
with increasing  $Q^2$.
Within the kinematical conditions of JLAB, 
the radiative correction is at the level of
several per cents or smaller if the hard cut on missing mass (or missing energy)
is used.

%
%
%

\section{Discussion and Conclusion}

In this paper we calculated radiative corrections to observable quantities in elastic electron-proton
scattering where polarization of the final proton is measured. Observable cross
section of this process has to include QED loop effects and contributions of
radiation of real photons and electron-positron pair creation
from leptonic line. In this paper the method of
structure functions is applied for this calculation. Within this approach it is
possible to calculate the contributions of leading and next-to-leading order in
all order of perturbation theory. Obtained explicit formulae are free from
infrared divergence and can be used for straightforward numerical analysis. This
numerical analysis was done for the kinematical condition of current and future
experiments at JLAB. The concrete values of radiative correction factors were
calculated. It was shown that radiative correction to observable ratio is at
the per cent level.

We note that the problem was solved for the case when kinematical variable $Q^2$ is
reconstructed via electron momentum measured. Another way is possible
for which
this variable is calculated using the measurement of final proton momentum.
This case requires another treatment, which will be done
elsewhere. Also the present calculation does not include effects due to
two-photon coupling to the proton.

The target considered in this paper is proton, however the results can be
straightforwardly generalized to the case when a nuclear target is used instead.
In this case the effects of Fermi motion and finite momentum of spectator
 nucleon system have to be taken into account.

\section*{Acknowledgements}
We thank our colleagues at Jefferson Lab for useful discussions.
We thank the US Department of Energy for support under contract
DE-AC05-84ER40150. Work of NM was in addition supported by Rutgers
University
through NSF grant PHY 9803860 and by Ukrainian DFFD.

\end{document}